\input phyzzx
\rightline {RHBNC-TH-96-2.}
\title {SYMPLECTIC EMBEDDINGS AND SPECIAL K\"AHLER GEOMETRY OF  $CP(n-1,1)$}
\author{W. A. Sabra\foot{e-mail: uhap012@vax.rhbnc.ac.uk}}
\address{Royal Holloway and Bedford New College,\break
University of London, Egham Surrey, \break
TW20 OEX, United Kingdom.}
\abstract{The embedding of the isometry group of
the coset spaces $\textstyle SU(1,n)\over \textstyle U(1)\times SU(n)$
in  $Sp(2n+2,{\hbox{\bf R}})$ is discussed.
The knowledge of such embedding provides a tool for the determination of the
holomorphic
prepotential characterizing the special geometry of these manifolds and
necessary in the
superconformal tensor calculus of $N=2$ supergravity. It is demonstrated that
there exists certain embeddings for
which the homogeneous prepotential does not exist. Whether a
holomorphic function exists or not, the dependence of the gauge kinetic terms
on the scalars characterizing these coset in $N=2$ supergravity theory
can be determined from the knowledge of the corresponding embedding,
\`a la Gaillard and Zumino. Our results are used to study some of the duality
symmetries of
heterotic compactifications of orbifolds with Wilson lines.}
\endpage
\REF\trieste{B. de Wit and A. Van Proeyen, {\it Special geometry and
symplectic transformation}, lectures delivered by A. Van Proeyen at the
String workshop on String theory. Trieste, April 1994, hep-th/9510186.}
\REF\lot{L. Andrianopoli, M. Bertolini, A. Ceresole, R. D'Auria,
S. Ferrara, P. Fre and T. Magri, hep-th/9605032.}
\REF\twelve{P. Candelas and X. de la Ossa,
{\it Nucl. Phys.} {\bf B355} (1991) 455;
P. Candelas,  X. de la Ossa, P. Green and L. Parkers,
{\it Nucl. Phys.} {\bf B359} (1991) 21.}
\REF\some{S. Ferrara, D. L\"ust and S. Theisen, {\it Phys. Lett.}
{\bf B242} (1990) 39.}
\REF\sw{N. Seiberg and E. Witten, {\it Nucl. Phys.} {\bf B426} (1994) 19.}
\REF\van{A. Ceresole, R. D'Auria, S. Ferrara and A. Van Proeyen,
{\it Nucl. Phys.} {\bf B444} (1995) 92.}
\REF\dw{B. de Wit, V. Kaplunovsky, J. Louis and D. L\"ust, {\it Nucl. Phys.}
{\bf B451} (1995) 53; I. Antoniadis, S. Ferrara, G. Gava, K. S. Narain and T.
R. Taylor,
{\it Nucl. Phys.} {\bf B447} (1995) 35.}
\REF\one{B. de Wit, P. G. Lauwers, R. Philippe, Su S. Q and A. Van
Proeyen, {\it Phys. Lett.}
{\bf B134} (1984) 37; B. de Wit and A. Van Proeyen, {\it Nucl. Phys.}{\bf B245}
(1984) 89.}
\REF\two{E. Cremmer, C. Kounnas, A. Van Proeyen, J. P. Derendinger,
S. Ferrara, B. de Wit and L. Girardello, {\it Nucl. Phys.} {\bf B250} (1985)
385.}
\REF\three{B. de Wit, P. G. Lauwers and A. Van Proeyen
{\it Nucl. Phys.} {\bf B255}
(1985) 569.}
\REF\four{E. Cremmer and A. Van Proeyen, {\it Class. Quantum Grav.}
{\bf 2} (1985) 445.}
\REF\five{S. Ferrara and A. Van Proeyen, {\it Class. Quantum Grav.}
{\bf 6} (1989) 124.}
\REF\six{A. Strominger, {\it Commun. Math. Phys.} {\bf 133}
(1990) 163.}
\REF\seven{L. Castellani, R. D'Auria and S. Ferrara, {\it Phys. Lett.}
{\bf B241} (1990) 57; R. D'Auria, S. Ferrara and P. Fr\`e, {\it Nucl. Phys.}
{\bf B359} (1991) {705}.}
\REF\GZ {M.K. Gaillard and B. Zumino, {\it Nucl. Phys.} {\bf B193}
(1981) 221.}

\REF\nine{S. Cecotti,
S. Ferrara, and L. Girardello, {\it Int. J. Mod. Phys.}
{\bf A4} (1989) 2475; {\it Phys. Lett.} {\bf  B213} (1988) 443}
\REF\ten{S. Ferrara and A. Strominger, in Strings
'89, ed. R. Arnowitt, R. Bryan, M. J. Duff, D. Nanopulos and C. N. Pope,
World Scientific, Singapore, (1990) 245.}
\REF\seiberg{N. Seiberg,{\it  Nucl. Phys.}{\bf B303} (1988) 286.}
\REF\eleven{L. Dixon, V. Kaplunovsky and J. Louis, {\it Nucl. Phys.}
{\bf B329} (1990) 27.}
\REF\fre{P. Fr\`e, P. Soriani, {\it Nucl
Phys.} {\bf B371} (1992) 659.}
\REF\eighteen{S. Ferrara, C. Kounnas,  D. L\"ust and F. Zwirner,
{\it Nucl. Phys.} {\bf B365} (1991) 431.}
\REF\short{W. A. Sabra, S. Thomas and N. Venegas, hep-th/9608075.}
\REF\fer{S. Ferrara, L. Girardello and M. Porrati,  {\it  Phys. Lett.} {\bf
B366} (1996) 155.}
\REF\orbifold{L. Dixon, J. A. Harvey, C. Vafa and E. Witten,
{\it Nucl. Phys.}{\bf B261} (1985) 678; {\bf B274} (1986) 285.}
\REF\nar{K. S. Narain, {\it Phys. Lett.} {\bf  B169} (1986) 41;
K. S. Narain, M. H. Sarmadi and E. Witten, {\it Nucl. Phys.} {\bf B279} (1987)
369.}
\REF\lust{G. L. Cardoso, D. L\"{u}st and T. Mohaupt, {\it  Nucl. Phys.} {\bf
B450} (1995) 115.}
\REF\luest{G. L. Cardoso, D. L\"{u}st and T. Mohaupt, {\it  Nucl. Phys.}
{\bf B432} (1994) 68.}
\REF\st{W. A. Sabra, {\it Mod. Phys. Letts.} {\bf A11} (1996) 1497.}

\chapter{INTRODUCTION}
In recent years, special geometry has emerged as an important structure
in the study of extended supergravity, superstrings and topological field
theories
\footnote*{for a review see [\trieste,\lot] and references therein.}.
The moduli space of $(2,2)$ superconformal field theories with central charge
$c=9$ exhibits special geometry. Special geometry played an important role
in the  analysis of Calabi-Yau threefolds in relation to mirror
and generalized duality symmetries [\twelve,\some]. More recently, special
geometry
provided a useful tool in the study of the quantum moduli space and obtaining
exact solutions of low energy effective actions for $N=2$
rigid [\sw] and local $N=2$ Yang Mills theories [\van,\dw].

The concept of special K\"ahler geometry first appeared in the
physics literature in the analysis of $N=2$ supergravity models
[\one,\two,\three].
There special K\"ahler manifolds are defined by the coupling of $n$
vector multiplets of the gauge sector of the theory to $N=2$ supergravity.
The lagrangian of the theory was derived using the
superconformal tensor calculus [\one]. In this method one starts with an action
invariant
under the $N=2$ superconformal group in four dimensions. Then with gauge fixing
conditions,
the resulting action is only invariant under super-Poincar\'e group as
required.
If we ignore the hypermultiplets, the theory contains
$(n+1)$ vector multiplets with scalar components $X^I,$
$(I=0,\cdots, n)$, where
the multiplet labelled by 0 corresponds to the graviphoton and
contains a scalar and a fermion to be gauge fixed in order
to break dilatations and the $U(1)$ symmetry of the
superconformal group and the a $R$-symmetry in the fermionic sector of the
theory.
It was found that the couplings can be described in terms of a prepotential $F$
which is a
holomorphic function of degree two in terms of the scalar fields, each of
weight one.
The physical scalar fields define an $n$-dimensional complex hypersurface
defined by the gauge fixing condition
$$i(X^I \bar F_I -F_I \bar X^I)=1.\eqn\co$$
Letting the $X^I$ be proportional to holomorphic sections $Z^I(z)$ of a
projective $(n+1)$-dimensional space [\seven], where $z$ is a set of $n$
complex coordinates, then the $z$ coordinates parametrize a K\"ahler space
with metric $g_{\alpha\bar\beta}=\partial_\alpha\partial_{\bar\beta}
K$, where $K$, the K\"ahler potential is expressed by
$$\eqalign{&K=-\log i\Big[Z^I\bar F_I(\bar Z)-
F_I(Z)\bar Z^I\Big]\cr
&=-log\Big(i<\Omega\vert\bar\Omega>\Big)=-\log i\pmatrix{Z^I&F_I
(Z)}
\pmatrix{\bf0&\bf1\cr -\bf1&\bf0}\pmatrix {\bar Z^I\cr \bar F_I(\bar Z)},\cr
& X^I=e^{K/2}Z^I,\qquad \bar X^I=e^{K/2}\bar Z^I.}\eqn\k$$
Special coordinates correspond to the choice
$$z^\alpha={X^\alpha\over X^0}; \qquad Z^0(z)=1,
\quad Z^\alpha(z)=z^\alpha.\eqn\madness$$
The homogeneous symmetric manifolds with the special
geometry structure
were classified in [\four] with appropriate holomorphic functions.

The gauge kinetic terms corresponding to the $(n+1)$ vector
multiplets  in an $N=2$ supergravity encoded in the matrix $\cal N$
is given by
$${\cal N}_{ij}=\bar F_{ij}+2i
{(Im F_{im})(Im F_{jn} )X^mX^n\over (Im F_{ab})X^aX^b}.\eqn\dd$$

The intrinsic definition of special K\"ahler geometry in
terms of symplectic bundles was later
given [\six] in connection with the geometry of the moduli of Calabi-Yau
spaces where special K\"ahler manifolds were associated with the moduli space
of
the K\"ahler or complex structure. In this approach the symplectic symmetry is
inherited from the symmetry of the homology cycles.
Also a special coordinate-independent description was given in [\seven],
where special geometry was obtained purely
from the constraints of the extended $N=2$ supersymmetry in the non-linear
sigma models associated with an arbitrary number $n$ of vector multiplets
of a four dimensional supergravity.
In this formalism, the underlying
symplectic $Sp(2n+2,{\hbox{\bf R}})$ symmetry is related to the duality
transformations
of the gauge sector as discussed in [\GZ].

In fact the relation between the above mentioned approaches
can be understood from the fact that on the same Calabi-Yau
manifold, both heterotic and type-II superstrings can be
compactified,
giving low-energy effective theories with $N=1$ and $N=2$
respectively. Therefore although the effective action of heterotic
compactification
does not have the $N=2$ structure, its moduli space must be
compatible
with the $N=2$ supersymmetry, giving it the special
structure [\nine,\ten,\seiberg]. The special geometry structure can also be
shown to be a consequence of the underlying $(2,2)$ world-sheet supersymmetry
[\eleven].

The only special K\"ahler manifolds which are of a direct
product form are given by [\five]
$$SK(n+1)={SU(1,1)\over U(1)}\times {SO(2,n)\over SO(2)
\times SO(n)}.\eqn\micha$$
These cosets are of fundamental importance in superstring theories, those with
$n=2, 4,$
describe the moduli spaces of $N=1$ heterotic string compactifications. They
also arise in
the study of heterotic
$N=2$ superstring
compactification,
where the $SU(1,1)\over U(1)$ represents the dilaton-axion vector multiplets
and the
other factor parametrize toroidal and possible Wilson-line moduli.

A method of constructing the
holomorphic prepotential for the manifolds $SK(n+1)$ has been given in
[\fre].  This construction is inspired by the method of Gaillard and Zumino
[\GZ] of
constructing the gauge couplings in an abelian gauge theory with
scalars parametrizing a coset space.
In this method, the duality transformations on the gauge fields
are parametrized by the embedding of the isometry group of the
coset space parametrized by the scalar fields in $Usp(n,n)$.
Such an embedding is crucial in fixing the lagrangian of the
theory which produces duality invariant equations of motion.

In the context of determining the holomorphic function encoding
the special geometry of the cosets $SK(n+1)$[\fre],
one introduces the symplectic section
$(X^\Lambda, F_\Lambda)$ and demands that it transforms as a vector under
the symplectic transformations induced
by the embedding of the isometry group of $SK(n+1)$ into
the symplectic group $Sp(2n+2,{\hbox{\bf R}}).$ These transformations are
then used to fix the relations between $F_\Lambda$ and $X^\Lambda$.
Clearly different embeddings
lead to different relations, and where an $F$ function exists, to different
$F$ functions.
In [\short] the results of [\fre] were extended to the other infinite
series of special manifolds, $CP_{n-1,1}={\textstyle SU(1,n)\over \textstyle
U(1)\times SU(n)}$.

In this paper we concentrate on the special K\"ahler manifolds $CP_{n-1,1},$
the construction of the corresponding kinetic terms for the scalars and
gauge fields in the corresponding $N=2$ supergravity.
We also discuss the relevance of our
formalism to the study of duality symmetries in orbifold compactifications
with Wilson lines.

This work is organized as follows. In section two we
review
the construction of [\short]. We give three  embeddings of the
isometry group $SU(1,n)$ into the symplectic group $Sp(2n+2,{\hbox{\bf R}}).$
This
construction is then applied to the simplest example ${SU(1,1)\over U(1)}$ and
we rederive
the familiar duality symmetry of this model. Section three contains an
explicit calculation for the gauge kinetic couplings of the $N=2$
supergravity theory whose scalars parametrize the coset $CP(n-1,1)$. This gives
an
emphasis on the importance of the embedding in the calculation of the
lagrangian in cases where a holomorphic function $F$ does not exist.
Moreover, $CP_{n-1,1}$ appear as submanifolds in heterotic string
compactifications on orbifolds with Wilson lines [\luest,\st].
The duality symmetry of these models can be
determined  from the study of the their mass spectrum.
In section four we will study
these duality transformations using the methods of special geometry.
In this formalism, the duality symmetries are much easier to analyze,
in particular for the cases with
more than one Wilson line. Section five contains a summary of our results and
conclusions.

\chapter{SYMPLECTIC EMBEDDINGS}
The isometry group of the cosets $CP_{n-1,1}$ is given by the group
$SU(1,n).$ An element of $SU(1, n)$ is represented by an $(n+1)\times (n+1)$
complex
matrix $M$ satisfying
$${M}^\dagger\eta {M}=\eta,\quad \det M={\bf1}, \eqn\mars$$ with
$\eta$ is the
constant diagonal metric with signature $(+, -,\cdots, -)$.
Decomposing the matrix $M$ into its real and imaginary part,
$${M}={U}+i{V},\eqn\dec$$ then the first relation in \mars\ implies
for the real $(n+1)\times (n+1)$ matrices ${ U}$ and ${V},$ the following
relations
$$\eqalign{{U}^t\eta {U}+{V}^t\eta {V}&=\eta,
\cr{ U}^t\eta {V}-{V}^t\eta {U}&=0.}\eqn\cybill$$
An element $\Omega$ of $Sp(2n+2,{\hbox{\bf R}})$ is a
$(2n+2)\times (2n+2)$ real matrix
satisfying
$${\Omega}^tL{\Omega}= L,\qquad
L=\pmatrix{{\bf 0}&{\bf 1}\cr -{\bf 1}&{\bf 0}},\eqn\pinhead$$
If we write $$\Omega=\pmatrix{{A}&{B}\cr {C}&{ D}}
\eqn\swansha$$
where the matrices $A$, $ B$ $C$ and $D$ are
$(n+1)\times (n+1)$ matrices, then in  terms of these block matrices,
\pinhead\ implies the following conditions
$${A}^t{C}-{C}^t{A}={\bf 0},\quad
{A}^t{D}-{C}^t{B}={\bf 1},\quad
{ B}^t{D}-{D}^t{B}={\bf 0}.\eqn\ero$$ An embedding of $SU(1,n)$
into the symplectic group $Sp(2n+2,{\hbox{\bf R}})$ is given by
$${A}={U}, \quad {C}=-\eta {V}, \quad {B}={V}\eta,
\quad {D}=\eta {U}\eta.\eqn\com$$

Consider the embedding ${\Omega}_e$ with matrix components
given in \com, and introduce the symplectic
section $(X^\Lambda, F_\Lambda)$ which
transforms as a vector under the symplectic transformations
induced by ${\Omega}_e$. These transformation rules can then be used
to determine the relation between $F_\Lambda$ and the
coordinates $X^\Lambda$.
In components, these transformations are given by
$$\eqalign{X &\rightarrow {U} X+{V}\eta\partial F,\cr
\partial F&\rightarrow -\eta {V} X+\eta {U}\eta \partial F,}
\eqn\tra$$
where $X$ and $\partial F$ are $(n+1)$-dimensional vectors
with components $X^\Lambda$ and $F_\Lambda$ respectively.
It is clear that the transformation relations \tra\  implies that $\partial F$
can be identified with
$i\eta X,$ in which case, a holomorphic prepotential $F$ exists
and is given,
in terms of the
coordinates $X$, by
$$F={i\over2}X^t\eta X.\eqn\Newyork$$
With the above relation, the complex vector $X$ transforms as
$$X\rightarrow ({U}+i{V})X={M}X,\eqn\dd$$
which implies that $X$ should be identified with the complex coordinates
which parametrize the $\textstyle SU(1, n)\over \textstyle U(1)\times SU(n)$
coset.
These complex coordinates
satisfy the following relation
$$\phi^\dagger\eta\phi={1}, \qquad\hbox{where}\quad
\phi=\left(\matrix{\phi^0\cr\vdots \cr\phi^{n+1}}\right),\eqn\nem$$
and are parametrized in terms of
unconstrained coordinates $z^\alpha$ by [\lust]
$$\phi^0={1\over \sqrt Y},\quad
\phi^j={z^\alpha\over \sqrt Y}, \quad \alpha=1, \cdots, n,\eqn\berlin$$
where $Y={1-\sum_\alpha z^\alpha\bar z^\alpha}.$
Here we identify $X$ with the complex vector ${1\over\sqrt2}\phi$.
The special coordinates in this case are given by
$z^\alpha$ and thus $Z^0=1,$ $Z^\alpha=z^\alpha,$ and the K\"ahler potential is
given by
$$K=-log(1-\sum_\alpha z^\alpha{\bar z}^\alpha).\eqn\so$$

A different embedding of $SU(1,n)$ into $Sp(2n+2,{\hbox{\bf R}})$ leads to
a different relation between $F_\Lambda$ and $X^\Lambda.$
In fact once an embedding $\Omega_e$ is specified,
then for all elements $S\in Sp(2n+2,{\hbox{\bf R}}),$
the matrix
$$\Omega'_e=S\Omega_e S^{-1},\eqn\ae$$ provides another embedding
with a corresponding symplectic section. As an example, consider the element
$$\eqalign{&S_1=\pmatrix{{\Sigma}&{\bf 0}\cr{\bf 0}&{\Sigma}},\quad
{\hbox{with}}
\cr &{\Sigma}=\pmatrix{{1\over\sqrt2}{\sigma}&{\bf 0}\cr{\bf 0}&{\bf 1}},
\qquad
\sigma=\pmatrix{1&1\cr 1&-1}.}\eqn\sis$$
Using \ae\ and \sis\ another
embedding of $SU(1,n)$ into $Sp(2n+2,{\hbox{\bf R}})$ can be obtained and is
given by
$$\Omega_e'=\pmatrix{{\Sigma}{U}{\Sigma}&{\Sigma}
{V}\eta{\Sigma}\cr -{\Sigma}\eta{V}{\Sigma}&
{\Sigma}\eta{U}\eta{\Sigma}}.\eqn\nst$$
For this embedding, we define a new section $(X',\partial F')$ which
transforms as a vector under the action of $\Omega'_e.$ Clearly,
the two sections $(X,\partial F)$ and $(X',\partial F')$ are related by
the following relations
$$X'={\Sigma}X,\qquad  {(\partial F)}'={\Sigma}\partial F.
\eqn\skeleton$$
These in componenets lead to the relations
$$\eqalign{X'^0&={1\over\sqrt2}(X^0+X^1),\cr
X'^1&={1\over\sqrt2}(X^0-X^1),\cr
X'^j&=X^j,\qquad j=2, \cdots, n\cr
F'_0&={1\over\sqrt2}(F_0+F_1)={i\over\sqrt2}(X^0-X^1)=iX'^1,\cr
F'_1&={1\over\sqrt2}(F_0-F_1)={i\over\sqrt2}(X^0+X^1)=iX'^0,\cr
F'_j&=F_j=-iX^j=-iX'^j.}\eqn\yael$$
{}From these relations, it can be easily seen that there exists a
holomorphic prepotential
$F'$ which can be expressed in terms of $X'$ by
$$F'={i}\Big(X'^0X'^1-{1\over2}\sum_{j=2}^n (X'^j)^2\Big).\eqn\nf$$
For this parametrization, we have
$$Z'^0=1,\quad Z'^1={1-z^1\over 1+z^1}, \qquad Z'^j={\sqrt 2 z^j\over
1+z^1},\eqn\sue$$
and the K\"ahler potential is given by
$$K=-\log (Z'^1+\bar Z'^1-\sum_{j}Z'^j\bar Z'^j).\eqn\lon$$
In general given two sections related by a symplectic transformations as
follows
$$\pmatrix{X'\cr\partial F'}=\pmatrix{{ X}&{ Y}\cr { Z}&{ T}}
\pmatrix{X\cr\partial F}.\eqn\leo$$
Then it can be shown that [\trieste] the relation between the
corresponding holomorphic functions is given by
$$F'={1\over2}\pmatrix{X&\partial F}\pmatrix{{Z}^t{X}&{Z}^t
{    Y}\cr { T}^t{ X}&{ T}^t{Y}}
\pmatrix{
X\cr\partial F}.\eqn\france$$
Moreover, it can be shown [\one] that a holomorphic prepotential $F'$ exists
such that
$$F'_\Lambda={\partial F'\over\partial X'^\Lambda},\eqn\susy$$
provided the mapping $X^\Lambda\rightarrow X'^\Lambda$ is invertible.

Using \Newyork,\france\ and the symplectic transformation
given in \sis, the expression of the holomorphic function
$F'$ given in \nf\ can be verified.

As a demonstration of the above calculations, we work out the
familiar example $SU(1,1)\over U(1)$ and derive the duality
symmetry action on the modulus of
this coset which is the special coordinate corresponding to the
two embedding described above. This example appears in string theory
in the description of the moduli spaces of heterotic string compactification as
well as
the moduli space parametrized by the complex dilaton-axion field.

First, let us define the special coordinates
$\textstyle t={\textstyle X^1\over\textstyle X^0}.$
Eq. \tra\ give the following transformation
for the coordinates $(X^0,X^1)$
$$\pmatrix{X^0\cr X^1}\rightarrow \pmatrix{z_1&\bar z_2\cr z_2&\bar z_1}
\pmatrix{X^0\cr X^1},\quad \vert z_1\vert^2-\vert z_2\vert^2=1.\eqn\demons$$
This gives for the special coordinates the following transformation
$$t\rightarrow {z_2+\bar z_1t\over z_1+\bar z_2t}.\eqn\anvy$$
The K\"ahler potential in terms of $t$ is given by
$$K= -\log(1-t\bar t).\eqn\napalm$$

In the coordinate system $({X'}^0,{X'}^1)$ corresponding to the embedding
\nst\ for the $SU(1,1)$ case, define a new special coordinate
$\textstyle T={\textstyle {X'}^1\over\textstyle {X'}^0}.$
If we write
$$\eqalign{(z_1+z_2)&=(d+ic),\cr (z_1-z_2)&=(a-ib).}
\eqn\scatter$$
Then the condition $z_1\bar z_1-z_2\bar z_2=1,$
implies that  $ad-bc=1.$
Using \nst\ we obtain
the following embedding in $Sp(4,{\hbox{\bf R}}),$
$$\Omega'_{SU(1,1)}
=\pmatrix{d&0&c&0\cr 0&a&0&-b\cr b&0&a&0\cr 0&-c&0&d}.\eqn\li$$
This gives using \nf, the following transformations
$$\eqalign{&{X'}^0\rightarrow d{X'}^0+ ic{X'}^1,\cr
&{X'}^1\rightarrow -ib{X'}^0+ a{X'}^1,}\eqn\peel$$
{}From which we obtain the familiar $SL(2,{\hbox{\bf R}})$ transformation for
the $T$ moduli
$$T\rightarrow {aT-ib\over icT+d}.\eqn\red$$
It is clear that the above two formalisms are related  by the holomorphic
field redefinition
$$T={1-t\over 1+t}.\eqn\moses$$

In what follows we will discuss a certain embedding for which
a holomorphic prepotential does not exist [\van].
Remaining with the $SU(1,1)\over U(1)$ example, consider the matrix
$$\pmatrix{d&0&c&0\cr 0&d&0&c\cr b&0&a&0\cr 0&b&0&a},\eqn\soul$$
it is clear that (using $ad-bc=1$) it provides an embedding of $SU(1,1)$ in
$Sp(4,{\hbox{\bf R}}).$
The transformation of the corresponding symplectic
vector,$(X''^\Lambda,F''_\Lambda)$, gives
$$\eqalign{X''^0&\rightarrow dX''^0+cF''_0,\cr
X''^1&\rightarrow dX''^1+cF''_1,\cr
F''_0&\rightarrow bX''^0+aF''_0,\cr
F''_1&\rightarrow dX''^1+aF''_1,}\eqn\loj$$
which implies that $F''_0$ and $X''^0$ are respectively proportional to $F''_1$
and $X''^1,$ and
therefore a holomorphic prepotential does not exist.

For the $SU(1,n)$ case, the embedding for which an $F$ function does not exist
can be obtained from the embedding $\Omega_e$ using \ae\ for
$S=S_2\in Sp(2n+2,{\hbox{\bf R}})$ given by
$$\eqalign{&S_2=\pmatrix{{ X}&{ Y}\cr { Z}&{ T}},
\quad \hbox{with}\quad
{ X}=\pmatrix{{ x}&{\bf 0}\cr {\bf 0}&{\bf 1}},\quad
{ Y}=\pmatrix{{ y}&{\bf 0}\cr {\bf 0}&{\bf 0}},\cr &
{ Z}=\pmatrix{{ z}&{\bf 0}\cr {\bf 0}&{\bf 0}},\quad
{ T}=\pmatrix{{ t}&{\bf 0}\cr {\bf 0}&{\bf 1}},\cr
&{ x}={ t}={1\over\sqrt2}\pmatrix{1&1\cr 0&0}, \qquad
{ z}=-{ y}={1\over\sqrt2}\pmatrix{0&0\cr 1&-1}.}\eqn\dale$$
The new section $(X''^\Lambda, F''_\Lambda)$
corresponding to the embedding
$$\Omega''_e=S_2\Omega_eS_2^{-1},\eqn\think$$
can be expressed in terms of $(X^\Lambda, F_\Lambda)$ by
$$\eqalign{X''^0&={1\over\sqrt2}(X^0+X^1),\cr
X''^1&={1\over\sqrt2}(F^1-F^0)=-{i\over\sqrt2}(X^0+X^1),\cr
X''^j&=X_j,\qquad j=2, \cdots, n\cr
F''_0&={1\over\sqrt2}(F^0+F^1)={i\over\sqrt2}(X^0-X^1),\cr
F''_1&={1\over\sqrt2}(X^0-X^1),\cr
F''_j&=F_j=X^j, \qquad j=2, \cdots, n}.$$
Using \Newyork, \france\ and \dale, it can be shown that $F''=0$.
Notice that the mapping $X^\Lambda\rightarrow X''^\Lambda$ is not invertible.
\vfill\eject
\chapter{Gauge Kinetic Couplings in $N=2$ supergravity}
In this section we discuss the gauge kinetic couplings of $N=2$ supergravity in
the formalism
of the theory corresponding to the various  embeddings considered in the
previous section using the method of [\GZ].
We demonstrate that the same results can be obtained directly using
the method of special geometry where the formalism admits a holomorphic
prepotential.

The action for the gauge part of the theory which also has a set of scalars
spanning an $m$-dimensional manifold can be written
using the notation of [\trieste] as
$${\cal L}={1\over4}(Im{\cal N}_{IJ}){\cal F}^I_{\mu\nu}
{\cal F}^{\mu\nu J}-{i\over8}(Re{\cal N}_{IJ})\epsilon^{\mu\nu\rho\sigma}{\cal
F}_{\mu\nu}^I
{\cal F}_{\rho\sigma}^J={1\over2}Im\Big({\cal N}_{IJ}{\cal F}^{+I}_{\mu\nu}
{\cal F}^{+\mu\nu J}\Big),\eqn\boil$$
where $I, J$ label the gauge fields, and ${\cal F}_{\mu\nu}^I$
denotes the field strength. Here $Im{\cal N}_{IJ}$ and $Re{\cal N}_{IJ}$
are, respectively, field dependent generalization of the inverse squared
coupling constant and the $\theta$-angle in standard gauge theories.

The Bianchi identities and equations of motion are given, respectively, by
$$\eqalign{&\partial^\mu Im{\cal F}^{+I}_{\mu\nu}=0,\cr
&\partial_\mu Im{\cal G}^{\mu\nu}_{+I}=0,}\eqn\swe$$
where
$$\eqalign{&{\cal G}_{+I}^{\mu\nu}\equiv 2i{\partial{\cal L}\over\partial{\cal
F}^{+I}_{\mu\nu}}
={\cal N}_{IJ}{\cal F}^{+I\mu\nu},\cr
&{\cal G}_{-I}^{\mu\nu}\equiv -2i{\partial{\cal L}\over\partial{\cal
F}^{-I}_{\mu\nu}}
=\bar{\cal N}_{IJ}{\cal F}^{-I\mu\nu},}\eqn\chains$$
The set of equations in \swe\ is invariant under $GL(2n,{\hbox{\bf R}})$
transformations:
$$\pmatrix{\hat{\cal F}^+\cr \hat{\cal G}_+}=\pmatrix {{ A}&{ B}\cr {C}&{ D}}
\pmatrix{{\cal F}^+\cr {\cal G}_+}.\eqn\sleep$$
To preserve the relations \chains\ under the action of the above transformation
implies that
the gauge kinetic coupling matrix must transform as
$$\hat{\cal N}=({ C}+{ D}{\cal N})({ A}+{ B}{\cal N})^{-1}.\eqn\basic$$
Moreover, the fact that the matrix $\cal N$ is symmetric
restricts the group of general linear transformations to
$Sp(2n,{\hbox{\bf R}}).$

The construction of the lagrangian \boil\ amounts to the determination of the
dependence of
$\cal N$ on the
scalar fields. Such dependence should produce the transformation law
of $\cal N$ defined in \basic.
Suppose that the scalar fields are valued in the coset space ${G\over H}$. The
part of the lagrangian involving the scalars only is invariant
under the isometry group of the coset. The isometry group action on the
gauge part of
the theory must correspond to a duality transformation as given in \sleep.
Therefore one has to imbed the isometry group into the group $Sp(2n,{\hbox{\bf
R}})$ or
$Usp(n, n)$ [\GZ, \lot].
An element of $Usp(n,n),$
we call $\cal S$, is a complex matrix which satisfies the symplectic
condition \pinhead\ together with the condition
$${\cal S}^\dagger{ J}{\cal S}={ J},\quad { J}=\pmatrix{{\bf 1}&{\bf0}\cr
{\bf0}
&-{\bf 1}}.\eqn\affair$$

An element of $Usp(n,n)$ can be given in terms of $\Omega\in Sp(2n,{\hbox{\bf
R}})$
defined in \pinhead\ by
$${\cal S}={\cal C}\Omega{\cal C}^{-1},\qquad {\cal C}=
{1\over\sqrt2}\pmatrix{{\bf 1}&i{\bf 1}\cr {\bf 1}&-i{\bf 1}}.\eqn\tr$$
This gives, for instance, for $\Omega_e\in Sp(2n,{\hbox{\bf R}})$ defined in
\com\ an element
in $Usp(n,n)$ given by
$${\cal S}={1\over2}\pmatrix{ U-i V\eta-i\eta V+\eta  U\eta&
 U+i V\eta-i\eta V-\eta  U\eta\cr  U-
i V\eta+i\eta V-\eta  U\eta,& U+i V\eta+i\eta V+\eta  U\eta
}.\eqn\for$$
If we associate to the coset representative
of $G\over H$
an element in $Usp(n,n)$ given by
$$\pmatrix {{ a}&{ b}^*\cr { b}&{ a}^*}.\eqn\tammara$$
It can then be shown that the matrix $\cal N$ is given by [\lot, \GZ],
$${\cal N}=i({ a}^\dagger+{ b}^\dagger)^{-1}
({ a}^\dagger-{ b}^\dagger).\eqn\swan$$

Let us consider the construction of the gauge kinetic matrix
for the simplest case where we have one physical scalar field
parametrizing the coset
${SU(1,1)\over U(1)}.$
An element of $SU(1,1)$ is given by the following matrix
$${ M}_1=\pmatrix{z_1 &\bar z_2\cr z_2&\bar z_1},\quad
\vert\bar z_1\vert^2-\vert\bar z_2\vert^2=1.\eqn\rep$$
Using \com\ an embedding of ${ M}_1$ in $Sp(4,{\hbox{\bf R}})$
can be given by
$${\Omega}_1=\pmatrix{u_1&u_2&v_1&v_2\cr u_2&u_1&v_2&v_1\cr
-v_1&v_2&u_1&-u_2\cr
v_2&-v_1&-u_2&u_1}.\eqn\hellbound$$
where we have decomposed ${ M}_1$ into its real and imaginary part
$$z_1=u_1+iv_1,\qquad  z_2=u_2+iv_2.\eqn\sophie$$
Eq. \tr\ or \for\ gives for ${ M}_1$, the following
embedding in $Usp(2,2)$
$${\cal S}_1=
\pmatrix{\bar z_1&0&0&z_2\cr 0&\bar z_1&z_2&0\cr
0&\bar z_2&z_1&0\cr \bar z_2&0&0&z_1}.\eqn\diana$$
A coset representative of $SU(1,1)\over U(1)$ can be given by
$${ W}_1=\pmatrix{\phi^0 &\bar\phi^1\cr \phi^1&\phi^0},\eqn\re$$
where $\phi_0$ and $\phi_1$ are as given in \berlin. The corresponding element
for
the coset representative in $Usp(n,n)$ is given by
$${\cal W}_{1}=
\pmatrix{\phi^0&0&0&\phi^1\cr 0&\phi^0&\phi^1&0\cr
0&\bar\phi^1&\phi^0&0\cr \bar\phi^1&0&0&\phi^0}.\eqn\diana$$

Using the expression \swan,
with $ a$ and $ b$ given by
$${ a}=\pmatrix{\phi^0&0\cr 0&\phi^0},\qquad
{ b}=\pmatrix{0&\bar\phi^1\cr \bar\phi^1&0},\eqn\hyde$$
we obtain for the gauge kinetic matrix
$$\eqalign{{\cal N}&=i\pmatrix{\phi^0&\phi^1\cr \phi^1&\phi^0}^{-1}
\pmatrix{\phi^0&-\phi^1\cr -\phi^1&\phi^0}\cr
&={i\over {(\phi^0)^2-(\phi^1)^2}}
\pmatrix{(\phi^0)^2+(\phi^1)^2&-2\phi^0\phi^1\cr
-2\phi^0\phi^1&(\phi^0)^2+(\phi^1)^2}\cr
&={i\over {(X^0)^2-(X^1)^2}}
\pmatrix{(X^0)^2+(X^1)^2&-2X^0X^1\cr -2X^0X^1&(X^0)^2+(X^1)^2}\cr
&={i\over {(1-t^2)}}
\pmatrix{1+t^2&-2t\cr -2t&1+t^2}.}\eqn\fish$$

Another embedding of ${ M}_1$ in $Sp(4,{\hbox{\bf R}})$
can be obtained using \nst\ for the $SU(1,1)$ case,  this is given by
$${\Omega}'_1=
\pmatrix{u_1+u_2&0&v_1+v_2&0\cr 0&u_1-u_2&0&v_1-v_2\cr v_2-v_1&0&u_1-u_2&0
\cr
0&-(v_1+v_2)&0&u_1+u_2},\eqn\hellbound$$
and the corresponding embedding in $Usp(2,2)$ is given by
$${\cal S}'_1=
\pmatrix{\bar z_1&0&z_2&0\cr
0&\bar z_1&0&-z_2\cr
\bar z_2&0&z_1&0\cr
0&-\bar z_2&0&z_1},
\eqn\darkne$$

The corresponding element for the coset representative in $Usp(2,2)$ is given
by
$${\cal W'}_1=\pmatrix{\phi^0&0&\phi^1&0\cr
0&\phi^0&0&-\phi^1\cr
\bar\phi^1&0&\phi^0&0\cr
0&-\bar\phi^1&0&\phi^0}.
\eqn\darkness$$
This gives using \swan\
the following gauge kinetic matrix
$${\cal N}'=i\pmatrix{{\textstyle X'^1\over\textstyle X'^0}&0\cr 0&{\textstyle
X'^0\over
\textstyle X'^1}}=
i\pmatrix{{T}&0\cr 0&{\textstyle 1\over\textstyle T}}=
i\pmatrix{{1-t\over 1+t}&0\cr 0&{1+t\over 1-t}}.\eqn\fedup$$

The embedding given in \soul, for which a holomorphic
function does not exist, give for
an element of $SU(1,1)$
an embedding in $Usp(2,2)$
given by
$${\cal S}''_1=
\pmatrix{\bar z_1&0&z_2&0\cr 0&\bar z_1&0& z_2\cr
\bar z_2&0&z_1&0\cr 0&\bar z_2&0&z_1}.\eqn\america$$
In this case, the coset representative of $SU(1,1)\over U(1)$
has the following embedding in $Usp(2,2)$
$${\cal W''}_{1}=\pmatrix{\phi^0&0&\phi^1&0\cr
0&\phi^0&0&\phi^1\cr
\bar\phi^1&0&\phi^0&0\cr
0&\bar\phi^1&0&\phi^0},
\eqn\dar$$
and the gauge kinetic terms for this embedding are given by
$${\cal N}''=i\pmatrix{{\textstyle X'^1\over\textstyle X'^0}&0\cr 0&{\textstyle
X'^1\over
\textstyle X'^0}}=i\pmatrix{T&0\cr 0&T}={i\over (1+t)}\pmatrix{1-t&0\cr
0&1-t}.\eqn\berger$$
Notice that the matrix ${\cal N}$ for the case where the $F$ function does not
exist can be
determined from the knowledge of the corresponding embedding.

As another example, consider the coset
$\textstyle SU(1,2)\over\textstyle U(1)\times SU(2)$, parametrized by two
scalar
fields $Z^1$ and $Z^2$. A coset representative is given by
$${ W}_2=\pmatrix{\phi^0&\bar\phi^1&\bar\phi^2\cr \phi^1& a& b\cr\phi^2& c&
d}.\eqn\mare$$
This gives for the embedding \com, a corresponding element in $Usp(3,3)$ given
by
$${\cal W}_{2}=\pmatrix{\phi^0&0&0&0&\phi^1&\phi^2\cr
0&a&b&\phi^1&0&0\cr 0&c&d&\phi^2&0&0\cr
0&\bar\phi^1&\bar\phi^2&\phi^0&0&0\cr \bar\phi^1&0&0&0&\bar a&\bar b\cr
\bar\phi^2&0&0&0&\bar c&\bar d},\eqn\karen$$
which produces the following gauge kinetic matrix
$${\cal N}=i\pmatrix{\phi^0&\phi^1&\phi^2\cr
\phi^1&\bar a&\bar c\cr\phi^2&\bar b&\bar d}^{-1}
\pmatrix{\phi^0&-\phi^1&-\phi^2\cr
-\phi^1&\bar a&\bar c\cr-\phi^2&\bar b&\bar d},\eqn\parrot$$
which in components gives
$$\eqalign{&{\cal N}_{00}={i\over\delta}\Big(
(\bar a\bar d-\bar b\bar c)\phi^0-(\bar b+\bar c)\phi^1\phi^2+
\bar a(\phi^2)^2+\bar d(\phi^1)^2\Big),\cr
&{\cal N}_{01}={-2i(\bar a\bar d-\bar b\bar c)\phi^1\over\delta},\cr
&{\cal N}_{02}={-2i(\bar a\bar d-\bar b\bar c)\phi^2\over\delta},\cr
&{\cal N}_{10}={i(2\bar c\phi^0\phi^2-2\bar d\phi^1\phi^0)\over\delta},\cr
&{\cal N}_{11}={i\over\delta}
\Big((\bar a\bar d-\bar b\bar c)\phi^0+(\bar b-\bar c)\phi^1\phi^2-
\bar a(\phi^2)^2+\bar d(\phi^1)^2\Big),\cr
&{\cal N}_{12}={i\over\delta}\Big(-2\bar c(\phi^2)^2+2\bar
d\phi^1\phi^2\Big),\cr
&{\cal N}_{20}={i\over\delta}\Big(2\bar b\phi^0\phi^1-2\bar a\phi^0\phi^2\Big),
\cr
&{\cal N}_{21}={i\over\delta}\Big(2\bar a\phi^1\phi^2-2\bar
b(\phi^1)^2\Big),\cr
&{\cal N}_{22}={i\over\delta}
\Big((\bar a\bar d-\bar b\bar c)\phi^0-(\bar b-\bar c)\phi^1\phi^2+
\bar a(\phi^2)^2-\bar d(\phi^1)^2\Big),\cr
&\delta=(\bar a\bar d-\bar b\bar c)\phi^0+(\bar b+\bar c)\phi^1\phi^2
-\bar a(\phi^2)^2-\bar d(\phi^1)^2.}\eqn\yo$$
Using the following conditions
$$ad- bc=\phi^0,\quad
a\bar\phi^2-b\bar\phi^1=\bar\phi^2,\quad
d\bar\phi^1-c\bar\phi^2=\bar\phi^1,\eqn\suffocate$$
which arise from the fact that ${ W}_2$ is an element of $SU(1,2),$
eq. \yo\ gives
$$\eqalign{&{\cal N}_{00}=
i{1+(Z^1)^2+(Z^2)^2\over 1-(Z^1)^2-(Z^2)^2},\cr
&{\cal N}_{01}={\cal N}_{10}=
-2i{Z^1\over 1-(Z^1)^2-(Z^2)^2},\cr
&{\cal N}_{02}={\cal N}_{02}=-2i{Z^2
\over 1-(Z^1)^2-(Z^2)^2},\cr
&{\cal N}_{11}=
i{ 1+(Z^1)^2-(Z^2)^2\over 1-(Z^1)^2-(Z^2)^2},\cr
&{\cal N}_{12}={\cal N}_{21}=2i {Z^1Z^2\over 1
-(Z^1)^2-(Z^2)^2},\cr
&{\cal N}_{22}=i{ 1-(Z^1)^2+(Z^2)^2
\over 1-(Z^1)^2-(Z^2)^2}.}\eqn\done$$

Another embedding of the coset representative in $Usp(3,3)$ can be given using
\nst\ by
$${\cal W}'_2=\pmatrix{{\textstyle\phi^0+a\over\textstyle2}&
{\textstyle\phi^0-a\over\textstyle2}&
{\textstyle b\over\textstyle\sqrt2} &
\textstyle\bar\phi^1&
0&{\textstyle\bar\phi^2\over\textstyle\sqrt2}\cr
\textstyle\phi^0-a\over\textstyle2&\textstyle\phi^0+a\over\textstyle2&-
{\textstyle b\over\textstyle\sqrt2}&0&-
\textstyle\bar\phi^1&\textstyle\bar\phi^2\over\textstyle\sqrt2\cr
\textstyle c\over\textstyle\sqrt2 &-{\textstyle c\over\textstyle\sqrt2}&d&
\textstyle\bar\phi^2\over\textstyle\sqrt2&\textstyle\bar\phi^2\over\textstyle\sqrt2&0\cr
\textstyle\phi^1&0&\textstyle\phi^2\over\textstyle\sqrt2&{\textstyle\phi^0+\bar
a\over
\textstyle2}
&{\textstyle\phi^0-\bar a\over\textstyle2}&
{\textstyle\bar b\over\textstyle\sqrt2}\cr
0&-\textstyle\phi^1&{\textstyle\phi^2\over\textstyle\sqrt2}&
{\textstyle\phi^0-\bar a\over\textstyle2}&
{\textstyle\phi^0+\bar a\over\textstyle2}&
-{\textstyle\bar b\over\textstyle\sqrt2}\cr
{\textstyle\phi^2\over\textstyle\sqrt2}&{\textstyle\phi^2\over\textstyle\sqrt2}&0&
{\textstyle\bar c\over\textstyle\sqrt2}&
-{\textstyle\bar c\over\textstyle\sqrt2}&\bar d}.
\eqn\karen$$
The gauge couplings in this case are given by
$$\eqalign{&{\cal N}'_{00}=
i{(1-Z^1)^2\over 1-(Z^1)^2-(Z^2)^2},\cr
&{\cal N}'_{01}={\cal N}'_{10}=
i{(Z^2)^2\over 1-(Z^1)^2-(Z^2)^2},\cr
&{\cal N}'_{02}={\cal N}'_{02}=-i{\sqrt2}{Z^2(1-Z^1)
\over 1-(Z^1)^2-(Z^2)^2},\cr
&{\cal N}'_{11}=
i{ (1+Z^1)^2\over 1-(Z^1)^2-(Z^2)^2},\cr
&{\cal N}'_{12}={\cal N}'_{21}=-i{\sqrt2}{Z^2(1+Z^1)
\over 1-(Z^1)^2-(Z^2)^2},\cr
&{\cal N}'_{22}=i{ 1-(Z^1)^2+(Z^2)^2
\over 1-(Z^1)^2-(Z^2)^2}.}\eqn\don$$

Finally for the embedding \soul\ we get the following element for the coset
representative
in $Usp(3,3)$
$${\cal W}''_2=
\pmatrix{{\textstyle
\phi^0+a\over\textstyle2}&-i{\textstyle\phi^0-a\over\textstyle2}&
{\textstyle b\over\textstyle\sqrt2}&\textstyle\bar\phi^1&
0&{\textstyle\bar\phi^2\over\textstyle\sqrt2}\cr
i{\textstyle\phi^0-a\over\textstyle2}&{\textstyle\phi^0+a\over\textstyle2}&-i
{\textstyle
b\over\textstyle\sqrt2}&0&\textstyle\bar\phi^1&i{\textstyle\bar\phi^2
\over\textstyle\sqrt2}\cr
{\textstyle c\over\textstyle\sqrt2}&i{\textstyle c\over\textstyle\sqrt2}&d&
{\textstyle\bar\phi^2\over\textstyle\sqrt2}&i{\textstyle\bar\phi^2\over\textstyle\sqrt2}&0\cr
\textstyle\phi^1&0&{\textstyle\phi^2\over\textstyle\sqrt2}&{\textstyle\phi^0+\bar a
\over\textstyle2}&i{\textstyle\phi^0-\bar a\over\textstyle2}&
{\textstyle\bar b\over\textstyle\sqrt2}\cr
0&\textstyle\phi^1&-i{\textstyle\phi^2\over\textstyle\sqrt2}&-i{\textstyle\phi^0-\bar a
\over\textstyle2}&{\textstyle\phi^0+\bar a\over\textstyle2}&
i{\textstyle\bar b\over\textstyle\sqrt2}\cr
{\textstyle\phi^2\over\textstyle\sqrt2}&-i{\textstyle\phi^2\over\textstyle\sqrt2}&0&
{\textstyle\bar c\over\textstyle\sqrt2}&-i{\textstyle\bar
c\over\textstyle\sqrt2}&\bar d}
\eqn\caren$$
and the gauge couplings in this case are given by
$$\eqalign{&{\cal N}''_{00}=i{1-(Z^1)^2+(Z^2)^2\over
(1+Z^1)^2},\cr
&{\cal N}''_{01}={\cal N}''_{10}=
-{(Z^2)^2\over 1+(Z^1)^2},\cr
&{\cal N}''_{02}={\cal N}''_{20}=-i{\sqrt2}{Z^2
\over 1+(Z^1)},\cr
&{\cal N}''_{11}=
i{ 1-(Z^1)^2-(Z^2)^2\over (1+Z^1)^2},\cr
&{\cal N}''_{12}={\cal N}'_{21}={\sqrt2}{Z^2
\over 1+Z^1},\cr
&{\cal N}'_{22}=i.}\eqn\doe$$

In general we represent the cosets $\textstyle SU(1,n)\over \textstyle
SU(n)\times U(1)$
by the element
$${ W}_n=\pmatrix{\phi^0&{\cal X}\cr{\cal  X}^\dagger&{\cal Y}},\eqn\topol$$
where
$${\cal X}=\pmatrix{\bar\phi^1&\bar\phi^2&\cdots&\bar\phi^n}\eqn\act$$
and ${\cal Y}$ is an $(n\times n)$ complex matrix satisfying the conditions
$$\eqalign{&{\cal Y}^\dagger {\cal Y}={\bf 1}+{\cal X}^\dagger{\cal X},\cr
&\phi^0 {\cal X}-{\cal X}{\cal Y}={\bf 0},\cr &
\det { W}_n={\bf 1}.}\eqn\tues$$
The embedding of $W_n$ in $Usp(n+1, n+1)$ is given by
$${\cal W}_n=\pmatrix{\phi^0&{\bf0}&{\bf0}&\bar{\cal X}\cr {\bf0}&{\cal
Y}&{\cal X}^\dagger&{\bf0}\cr {\bf0}&{\cal X}&\phi^0&{\bf0}\cr {\cal
X}^t&{\bf0}&{\bf0}&\bar{\cal Y}}.\eqn\sop$$

The other two embeddings ${\cal W}'_{n}$ and ${\cal W}''_{n}$ can be
obtained from ${\cal W}_{n}$ by the following relations
$$\eqalign{&{\cal W}'_{n}={\cal C}_1{\cal W}_{n}{\cal C}_1^{-1},\qquad
{\cal W}''_{n}={\cal C}_2{\cal W}_{n}{\cal C}_2^{-1},\cr &
{\cal C}_1={\cal C}S_1{\cal C}^{-1},\qquad
{\cal C}_2={\cal C}S_2{\cal C}^{-1}.}\eqn\mea$$
where $\cal C$, $S_1$ and $S_2$ are given, respectively, given in \tr,
\sis\ and \dale.

Having determined the form of the gauge kinetic matrix in the various
embeddings considered in this paper, we demonstrate how our results
can be obtained using the formalism of special geometry.
For $\textstyle SU(1,n)\over \textstyle SU(n)\times U(1)$,
the holomorphic function in the basis $X$ and $X'$ where shown to
be given by \Newyork\ and \nf.
Using the expressions of $F$ and $F'$ for the $n=1$ case, one can reproduce
the expressions for $\cal N$ and ${\cal N}'$ in \fish\ and \fedup\
respectively.
The reader could verify for herself that substituting
the expressions for the holomorphic functions for the various cosets
produces the results obtained by using the method of [\GZ].

The matrix ${\cal N}'$ and ${\cal N}''$ in the basis $X'$ and $X''$ can also be
obtained
from $\cal N$ by performing a symplectic transformation which
connects the section $(X^\Lambda, F_\Lambda)$ with
$(X'^\Lambda, F'_\Lambda)$ and $(X''^\Lambda, F''_\Lambda)$. This gives
$$\eqalign{&{\cal N'}=\Sigma{\cal N}\Sigma,\cr
&{\cal N}''=({ Z}+{ T}{\cal N})({ X}+{ Y}{\cal N})^{-1}.}\eqn\hoffman$$
\vfill\eject
\chapter{Duality symmetries}
In this section our analysis of the cosets
$\textstyle SU(1,n)\over \textstyle U(1)\times SU(n)$ is used to study
a subgroup of the duality
symmetry in heterotic string theories compactified on orbifolds with Wilson
lines
[\orbifold,\luest].
In toroidal compactifications [\nar], one has a set of scalar fields, the
moduli, which are encoded in the metric of the lattice
defining the torus and a possible antisymmetric tensor and Wilson lines.
The moduli space of toroidal compactification [\nar] is given (locally)
by the coset space ${SO(d+16,d)\over SO(d+16,d)\times
SO(d)},$
where $d$ is the dimension of the torus upon which the theory is
compactified
and the factor $16$ comes from the inclusion of Wilson lines.
In orbifold models, the twist
freezes some of the moduli and thus the moduli
spaces of orbifolds have smaller dimensions than those of their corresponding
toroidal compactifications.
It can be demonstrated [\luest] that the moduli
spaces of orbifolds can be determined from the knowledge
of the eigenvalues of the twist and their multiplicities.
The moduli space of the orbifold (without Wilson lines)
are parametrized by the $T$ moduli corresponding to the K\"ahler
deformations  and the $U$ moduli which correspond to the
deformations of the complex structure.
Both moduli spaces are given by a special K\"ahler manifold.
The $U$ moduli space is  described by the coset
$\Big[{\textstyle SU(1,1)\over\textstyle U(1)}\Big]$, and except
for the ${\bf Z}_3$ orbifold, whose $T$ moduli space is given by
${\textstyle{SU(3,3)}\over \textstyle{SU(3)\times SU(3)\times U(1)}},$
the $T$ moduli spaces for all  symmetric orbifolds yielding $N=1$ space-time
supersymmetry are given by the special K\"ahler manifolds.
$${SK}(n+1)={SU(1,1)\over U(1)}\times {SO(n,2)\over SO(n)
\times SO(2)},\qquad n=2,4.$$

Duality symmetries are those
discrete automorphisms of the moduli space which leave
the underlying conformal field theory invariant. One method of determining
the duality symmetry is to study the
mass spectrum of the theory which depends on the moduli fields
and quantum numbers. Duality transformations are identified by
those transformations on the moduli fields and quantum numbers which
leaves the spectrum invariant.

In what follows we shall analyze the duality symmetries of the cosets
$CP(n-1,1)$ which describe a submanifold in the moduli
space of factorizable orbifolds with Wilson lines [\luest, \st].
The examples for $n=1$ and $n=2$ were studied in details in [\st],
relying on the mass
spectrum of these cosets. Here, we shall reproduce these results using
the methods of special geometry, as well as the duality symmetry
and its action on the moduli for any value of $n$.
Clearly the number $n,$ related to the number of Wilson lines,
is not arbitrary and is constrained by modular invariance.

The mass formula for any special K\"ahler manifold was provided in
[\eighteen]. Once a section $(X, \partial F)$ is specified, based on symmetry
arguments,
the mass formula can be given by
$$\vert m\vert^2=\vert PX+Q\partial F\vert^2.\eqn\m$$
where $(P, Q)$ is a vector encoding the windings and momenta.
Under the target space duality group $\Gamma$,
the vector $(X, \partial F)$ transforms
by
$$\pmatrix{X\cr \partial F}\rightarrow \pmatrix{{ A}'&{ B}'\cr { C}'&
{ D}'}
\pmatrix{X\cr \partial F},\eqn\liz $$
where the matrix $\pmatrix{{ A}'&{ B}'\cr { C}'&
{ D}'}$ is the embedding of the duality
group in $Sp(2n+2,{\hbox{\bf Z}}).$
For $\vert m\vert^2$ to be invariant under the duality transformations,
$(P, Q)$ should transform as follows
$$\pmatrix {P\cr Q}\rightarrow\pmatrix{{ D}'&-{ C}'\cr -{ B}'&{ A}'}
\pmatrix{P\cr Q}.\eqn\mary$$

For $\Gamma=SU(1,n)$ and for the embedding defined in \com\ where the
holomorphic prepotential
given by \Newyork\ we have
$$\eqalign{&\pmatrix{X\cr \partial F}\rightarrow
\pmatrix{{ U}'&{ V}'\eta\cr -\eta { V}'&\eta{ U}'\eta}
\pmatrix{X\cr\partial F},\cr &\pmatrix{P\cr Q}\rightarrow
\pmatrix{\eta { U}'\eta&\eta{ V}'\cr -{ V}'\eta&{ U}'}\pmatrix{P\cr
Q},}\eqn\l$$
with ${ U}'$ and ${ V}'$ are integer valued matrices. Using the form of the
holomorphic prepotential in \Newyork, the mass formula can be written as
$$\vert m\vert^2=\vert P'^tX\vert^2,\quad \hbox{with}\quad
P'=\pmatrix{P'_0=P_0+iQ_0\cr
P'_i=P_i-iQ_i}.\eqn\house$$
Under the $SU(1,n)$ transformation
$$X\rightarrow M X,\qquad M\in SU(1, n),\eqn\man$$
it can be easily shown that the following transformations
for the quantum numbers hold,
$$\pmatrix{P'_0\cr -P'_i}\rightarrow M^*\pmatrix{P'_0\cr -P'_i}.\eqn
\he$$

We now compare these calculations with those of string compactifications.
First consider the simplest example of $\Gamma=SU(1,1)$.
The mass formula is given by [\st]
$$\vert m\vert^2=2{\vert m_c-n_ct\vert^2\over (1-t\bar
t)}=u_1^\dagger\Xi_1u_1,$$
where $$\eqalign{&\Xi_1={2\over 1-t\bar t}
\pmatrix{1&-t\cr -\bar t&t\bar t},\quad u_1=\pmatrix{m_c\cr n_c},\cr
& m_c={1\over{2\sqrt2}u_1}(m_2-im_1U_0+2in_1u_1-2n_2u_1U_0),\cr
& n_c={1\over{2\sqrt2}u_1}(-m_2+im_1U_0+2in_1u_1-2n_2u_1U_0),}\eqn\death$$
where we have have set the moduli
$U_0=u_1+iu_2$, and performed the change of variables
$t={\textstyle{1-T'}\over\textstyle{1+T'}},$ with
$T'={\textstyle T\over\textstyle 2u_1}$ and $(n_1,n_2, m_1,m_2)$ are the
winding and momentum quantum numbers.
Notice that in this example, the coset $SU(1,1)\over U(1)$
appears as atruncation of the coset $SO(2,2)\over SO(2)\times SO(2)$ when the
moduli $U$ is fixed to a
constant value by twisting. That is  why $m_c$ and $n_c$ are not integer
valued.
This breaks the duality group $SU(1,1,{\hbox{\bf Z}})$ down to a subgroup which
depends on the
value of $U_0$. For more details the reader is referred to [\st].

If we define the action of the duality group on the quantum numbers by
$$u_1\rightarrow M_1^*u_1,\eqn\stut$$
then for $\vert m\vert^2$ in \death\ to be target space duality invariant, we
get
$$\Xi_1\rightarrow{{M_1^t}^{-1}}\Xi_1{M_1^*}^{-1}.\eqn\shine$$
{}From \shine\  one can extract the duality transformation for the
moduli $t$. This can be easily done by writing
$$\Xi_1=\pmatrix{X^0\cr -\bar X^1}\pmatrix{X^0&-X^1}.\eqn\ealing$$
where we have defined $\textstyle t={\textstyle X^1\over\textstyle X^0}.$
Then \shine\  together with the relation
$${M_1^\dagger}^{-1}=LM_1 L, \quad L=\pmatrix{1&0\cr 0&-1}. \eqn\mary$$
give
$$\pmatrix{X^0\cr X^1}\rightarrow M_1\pmatrix{X^0\cr X^1},\eqn\hob$$
a result which agrees with \man.

Next consider a less trivial example given by the coset $SU(1,2)\over
SU(2)\times U(1).$
The mass formula in this case
is given by [\st]
$$\eqalign{&\vert m\vert^2={2\vert m_c-n_ct-Q_c{\cal A}\vert^2\over
1-t\bar t-{\cal A}\bar{\cal A}}
=u_2^\dagger\Xi_2 u_2,\cr
&\Xi_2={2\over 1-t\bar t-{\cal A}\bar{\cal A}}
\pmatrix{1&-t&-{\cal A}\cr -\bar t&t\bar t&
\bar t{\cal A}\cr -{\cal A}&t\bar{\cal A}&{\cal A}\bar{\cal A}},\qquad
u_2=\pmatrix {n_c\cr m_c\cr Q_c}}\eqn\flowers$$
here $Q_c$ depends on the embedding of the twist in $E_8\times E_8$ [\luest]
and $\cal A$
represents the Wilson line moduli.
Again under the duality transformation which is given by a subgroup of
$SU(1,2,{\hbox{\bf Z}})$ and parametrized by $M_2$, we have
$$u_2\rightarrow M^*_2u_2,\eqn\south$$
and
$$\Xi_2\rightarrow{{M_2^t}^{-1}}\Xi_2{M_2^*}^{-1}.\eqn\shin$$
If we define the moduli fields by
$$t={X^1\over X^0}, \qquad {\cal A}={X^2\over X^0},\eqn\dry$$
 $\Xi_2$ can be rewritten as
$$\Xi_2=\pmatrix{X^0\cr -\bar X^1\cr -\bar X^2}
\pmatrix{X^0&-X^1&-X^2}.\eqn\liberal$$
and  \shin\ then gives
$$\pmatrix{X^0\cr X^1\cr X^2}\rightarrow
\Omega_2\pmatrix{X^0\cr X^1\cr X^2},\eqn\nuclear$$
which agrees with \man. From \nuclear\ one can easily read off
the duality transformations of the moduli, these are given by
$$\eqalign{&t\rightarrow {z_3+z_4t+z_5{\cal A}\over z_0+z_1t+z_2{\cal A}},
\cr &{\cal A}\rightarrow {z_6+z_7t+z_8{\cal A}\over z_0+z_1t+z_2{\cal A}}.}
\eqn\tanks$$
Therefore, for a model with $n-1$ Wilson line moduli, where the duality group
is a subgroup of
$SU(1,n, {\hbox{\bf Z}})$,  eq. \man\ together with the identification
$$t={X^1\over X^0}, \qquad {\cal A}^j={X^j\over X^0},\eqn\wet$$
gives the duality transformation for any number of moduli.

\chapter{Conclusions}
We have analyzed the special K\"ahler manifolds, the so-called minimal
coupling,
$\textstyle SU(1, n)\over \textstyle U(1)\times SU(n)$ with regard to
the construction of their prepotentials.
The prepotential is a holomorphic  function  of degree two, expressed
in terms of the scalar fields parametrizing these cosets, and is essential in
the superconformal tensor calculus of $N=2$ supergravity. The method employed
which relies on the embedding
of the isometry group $SU(1,n)$ into the symplectic group $Sp(2n+2,{\hbox{\bf
R}})$ provides
a powerful tool in calculating the lagrangian of the model.
There are certain embeddings for which
a holomorphic function does not exist. In these cases the knowledge of the
embedding is enough to determine the lagrangian of the vectors and scalars
irrespective of the fact whether an $F$ function exists or not.
\

The analysis of these
cosets apart from being relevant
to $N=2$ supergravity models, is also important for the study of a
subgroup of the duality symmetries in heterotic
string compactifications with Wilson lines.
The target space symmetries are much easier to analyze in terms of the
symplectic sections. In terms of the coordinates of the section, the duality
symmetries are linear and form a subgroup of the symplectic transformations.

Recently, in the analysis of physically interesting problems,
the formalism of the theory requires the non-existence of the holomorphic
function.
This was shown to be the case in the study of
perturbative corrections to
vector couplings in $N=2$ heterotic string vacua [\van,\dw]. The non-existence
of the holomorphic function also
provided a new mechanism in the study of supersymmetry breaking of
$N=2$ supersymmetry down to $N=1$ [\fer].
In this sense, our analysis will be relevant in the study of quantum
corrections of the coupling in the minimal coupling models as well as in the
study of
supersymmetry breaking.
\vfill\eject

\centerline{ACKNOWLEDGEMENT}
I would like to thank J. Figueroa-O'Farrill for a useful discussion.
This work is supported by PPARC.
\vfill\eject
\refout
\end